\documentclass[aps,prl,superscriptaddress,showpacs,floatfix,twocolumn]{revtex4}

\usepackage{graphicx}   
\usepackage{amssymb}
\usepackage{subfigure}
\usepackage{pslatex}

\newcommand{\sqrtsnn}{\sqrt{s_{_{NN}}}}
\def\mean#1{\ensuremath{\left<#1\right>}}

\begin{document}

\title{Common Suppression Pattern of $\eta$ and $\pi^0$ Mesons\\
at High Transverse Momentum in Au+Au Collisions at $\sqrtsnn$ = 200 GeV}

\newcommand{\abilene}{Abilene Christian University, Abilene, TX 79699, U.S.}
\newcommand{\acadsin}{Institute of Physics, Academia Sinica, Taipei 11529, Taiwan}
\newcommand{\banaras}{Department of Physics, Banaras Hindu University, Varanasi 221005, India}
\newcommand{\barc}{Bhabha Atomic Research Centre, Bombay 400 085, India}
\newcommand{\bnl}{Brookhaven National Laboratory, Upton, NY 11973-5000, U.S.}
\newcommand{\caucr}{University of California - Riverside, Riverside, CA 92521, U.S.}
\newcommand{\ciae}{China Institute of Atomic Energy (CIAE), Beijing, People's Republic of China}
\newcommand{\cns}{Center for Nuclear Study, Graduate School of Science, University of Tokyo, 7-3-1 Hongo, Bunkyo, Tokyo 113-0033, Japan}
\newcommand{\columbia}{Columbia University, New York, NY 10027 and Nevis Laboratories, Irvington, NY 10533, U.S.}
\newcommand{\dapnia}{Dapnia, CEA Saclay, F-91191, Gif-sur-Yvette, France}
\newcommand{\debrecen}{Debrecen University, H-4010 Debrecen, Egyetem t{\'e}r 1, Hungary}
\newcommand{\fsu}{Florida State University, Tallahassee, FL 32306, U.S.}
\newcommand{\gsu}{Georgia State University, Atlanta, GA 30303, U.S.}
\newcommand{\hiroshima}{Hiroshima University, Kagamiyama, Higashi-Hiroshima 739-8526, Japan}
\newcommand{\ihepprot}{IHEP Protvino, State Research Center of Russian Federation, Institute for High Energy Physics, Protvino, 142281, Russia}
\newcommand{\illuiuc}{University of Illinois at Urbana-Champaign, Urbana, IL 61801}
\newcommand{\isu}{Iowa State University, Ames, IA 50011, U.S.}
\newcommand{\jinrdubna}{Joint Institute for Nuclear Research, 141980 Dubna, Moscow Region, Russia}
\newcommand{\kaeri}{KAERI, Cyclotron Application Laboratory, Seoul, South Korea}
\newcommand{\kangnung}{Kangnung National University, Kangnung 210-702, South Korea}
\newcommand{\kek}{KEK, High Energy Accelerator Research Organization, Tsukuba, Ibaraki 305-0801, Japan}
\newcommand{\kfki}{KFKI Research Institute for Particle and Nuclear Physics of the Hungarian Academy of Sciences (MTA KFKI RMKI), H-1525 Budapest 114, POBox 49, Budapest, Hungary}
\newcommand{\korea}{Korea University, Seoul, 136-701, Korea}
\newcommand{\kurchatov}{Russian Research Center ``Kurchatov Institute", Moscow, Russia}
\newcommand{\kyoto}{Kyoto University, Kyoto 606-8502, Japan}
\newcommand{\labllr}{Laboratoire Leprince-Ringuet, Ecole Polytechnique, CNRS-IN2P3, Route de Saclay, F-91128, Palaiseau, France}
\newcommand{\lawllnl}{Lawrence Livermore National Laboratory, Livermore, CA 94550, U.S.}
\newcommand{\losalamos}{Los Alamos National Laboratory, Los Alamos, NM 87545, U.S.}
\newcommand{\lpc}{LPC, Universit{\'e} Blaise Pascal, CNRS-IN2P3, Clermont-Fd, 63177 Aubiere Cedex, France}
\newcommand{\lund}{Department of Physics, Lund University, Box 118, SE-221 00 Lund, Sweden}
\newcommand{\muenster}{Institut f\"ur Kernphysik, University of Muenster, D-48149 Muenster, Germany}
\newcommand{\myongji}{Myongji University, Yongin, Kyonggido 449-728, Korea}
\newcommand{\nagasaki}{Nagasaki Institute of Applied Science, Nagasaki-shi, Nagasaki 851-0193, Japan}
\newcommand{\newmex}{University of New Mexico, Albuquerque, NM 87131, U.S.}
\newcommand{\nmsu}{New Mexico State University, Las Cruces, NM 88003, U.S.}
\newcommand{\ornl}{Oak Ridge National Laboratory, Oak Ridge, TN 37831, U.S.}
\newcommand{\orsay}{IPN-Orsay, Universite Paris Sud, CNRS-IN2P3, BP1, F-91406, Orsay, France}
\newcommand{\pnpi}{PNPI, Petersburg Nuclear Physics Institute, Gatchina,  Leningrad region, 188300, Russia}
\newcommand{\riken}{RIKEN, The Institute of Physical and Chemical Research, Wako, Saitama 351-0198, Japan}
\newcommand{\rikjrbrc}{RIKEN BNL Research Center, Brookhaven National Laboratory, Upton, NY 11973-5000, U.S.}
\newcommand{\saispbstu}{Saint Petersburg State Polytechnic University, St. Petersburg, Russia}
\newcommand{\saopaulo}{Universidade de S{\~a}o Paulo, Instituto de F\'{\i}sica, Caixa Postal 66318, S{\~a}o Paulo CEP05315-970, Brazil}
\newcommand{\seoulnat}{System Electronics Laboratory, Seoul National University, Seoul, South Korea}
\newcommand{\stonybrkc}{Chemistry Department, Stony Brook University, SUNY, Stony Brook, NY 11794-3400, U.S.}
\newcommand{\stonycrkp}{Department of Physics and Astronomy, Stony Brook University, SUNY, Stony Brook, NY 11794, U.S.}
\newcommand{\subatech}{SUBATECH (Ecole des Mines de Nantes, CNRS-IN2P3, Universit{\'e} de Nantes) BP 20722 - 44307, Nantes, France}
\newcommand{\tenn}{University of Tennessee, Knoxville, TN 37996, U.S.}
\newcommand{\titech}{Department of Physics, Tokyo Institute of Technology, Oh-okayama, Meguro, Tokyo, 152-8551, Japan}
\newcommand{\tsukuba}{Institute of Physics, University of Tsukuba, Tsukuba, Ibaraki 305, Japan}
\newcommand{\vandy}{Vanderbilt University, Nashville, TN 37235, U.S.}
\newcommand{\waseda}{Waseda University, Advanced Research Institute for Science and Engineering, 17 Kikui-cho, Shinjuku-ku, Tokyo 162-0044, Japan}
\newcommand{\weizmann}{Weizmann Institute, Rehovot 76100, Israel}
\newcommand{\yonsei}{Yonsei University, IPAP, Seoul 120-749, Korea}
\affiliation{\abilene}
\affiliation{\acadsin}
\affiliation{\banaras}
\affiliation{\barc}
\affiliation{\bnl}
\affiliation{\caucr}
\affiliation{\ciae}
\affiliation{\cns}
\affiliation{\columbia}
\affiliation{\dapnia}
\affiliation{\debrecen}
\affiliation{\fsu}
\affiliation{\gsu}
\affiliation{\hiroshima}
\affiliation{\ihepprot}
\affiliation{\illuiuc}
\affiliation{\isu}
\affiliation{\jinrdubna}
\affiliation{\kaeri}
\affiliation{\kangnung}
\affiliation{\kek}
\affiliation{\kfki}
\affiliation{\korea}
\affiliation{\kurchatov}
\affiliation{\kyoto}
\affiliation{\labllr}
\affiliation{\lawllnl}
\affiliation{\losalamos}
\affiliation{\lpc}
\affiliation{\lund}
\affiliation{\muenster}
\affiliation{\myongji}
\affiliation{\nagasaki}
\affiliation{\newmex}
\affiliation{\nmsu}
\affiliation{\ornl}
\affiliation{\orsay}
\affiliation{\pnpi}
\affiliation{\riken}
\affiliation{\rikjrbrc}
\affiliation{\saispbstu}
\affiliation{\saopaulo}
\affiliation{\seoulnat}
\affiliation{\stonybrkc}
\affiliation{\stonycrkp}
\affiliation{\subatech}
\affiliation{\tenn}
\affiliation{\titech}
\affiliation{\tsukuba}
\affiliation{\vandy}
\affiliation{\waseda}
\affiliation{\weizmann}
\affiliation{\yonsei}
\author{S.S.~Adler}	\affiliation{\bnl}
\author{S.~Afanasiev}	\affiliation{\jinrdubna}
\author{C.~Aidala}	\affiliation{\bnl}
\author{N.N.~Ajitanand}	\affiliation{\stonybrkc}
\author{Y.~Akiba}	\affiliation{\kek} \affiliation{\riken}
\author{J.~Alexander}	\affiliation{\stonybrkc}
\author{R.~Amirikas}	\affiliation{\fsu}
\author{L.~Aphecetche}	\affiliation{\subatech}
\author{S.H.~Aronson}	\affiliation{\bnl}
\author{R.~Averbeck}	\affiliation{\stonycrkp}
% ...
\collaboration{PHENIX Collaboration} \noaffiliation

\date{\today}

\begin{abstract}
Inclusive transverse momentum spectra of $\eta$ mesons have been measured within 
$p_T$ = 2 -- 10~GeV$/c$ at mid-rapidity by the PHENIX experiment in Au+Au collisions 
at $\sqrt{s_{_{NN}}}$ = 200~GeV. In central Au+Au the $\eta$ yields are significantly 
suppressed compared to peripheral Au+Au, d+Au and p+p yields scaled by the 
corresponding number of nucleon-nucleon collisions. The magnitude, centrality and 
$p_{T}$ dependence of the suppression is common, within errors, for $\eta$ and 
$\pi^0$. The ratio of $\eta$ to $\pi^0$ spectra at high $p_T$ amounts to 
0.40 $< R_{\eta/\pi^{0}} <$ 0.48 for the three systems in agreement with the 
world average measured in hadronic and nuclear reactions and, at large scaled 
momentum, in $e^+e^-$ collisions.
\end{abstract}

\pacs{25.75.Dw}

\maketitle

The major motivation for the study of high energy
nucleus-nucleus (A+A) collisions is the opportunity to probe
strongly interacting matter at extremely high energy densities.
Of particular interest are energy densities well above the expected
transition from normal hadronic matter to a deconfined system
of quarks and gluons.  Lattice Quantum Chromodynamics (QCD)
calculations~\cite{latt} predict that this transition will occur
at a temperature of T $\approx$ 170 MeV $\approx 10^{12}$ K.
The formation of a quark-gluon plasma (QGP) in A+A reactions 
should manifest itself in a variety of experimental 
signatures~\cite{harris96}.  

At center-of-mass energies reached by the Relativistic Heavy Ion Collider 
(RHIC), arguably the most exciting experimental results so far 
are connected with the predicted ``jet quenching'' phenomenon~\cite{bjorken82,gyulassy90,bdmps}
due to energy loss of hard-scattered partons as they traverse the dense medium 
formed in the reaction. Since (leading) hadrons with $p_T>$ 4 GeV/$c$ at RHIC 
carry a large fraction of the momentum of the parent quark or gluon
($\mean{z}=p_{\ensuremath{\it hadron}}/p_{\ensuremath{\it parton}}\approx$ 
0.5 -- 0.7~\cite{ppg029,kretzer}), parton energy loss results in a significantly 
suppressed production of high-$p_T$ hadrons~\cite{gyulassy90}.
The inclusive spectra of high-$p_T$ neutral pions~\cite{ppg014,ppg054} 
and charged hadrons~\cite{star_hipt_200,phenix_hipt_200} 
in Au+Au at $\sqrt{s_{_{NN}}}$ = 200~GeV are indeed suppressed by as much as 
a factor of five compared to the corresponding 
yields in p+p~\cite{ppg024} and d+Au~\cite{ppg028,ppg044},
scaled by the number of incoherent nucleon-nucleon ($NN$) collisions. 
The centrality~\cite{wang03}, $p_T$~\cite{vitev_gyulassy,jeon_moore,eskola04} and 
center-of-mass energy~\cite{dde_hp04} dependences of the observed quenching 
are consistent with theoretical calculations of QCD energy loss
due to multiple gluon emission in a dense medium. 
Assuming a thermalized parton system, the magnitude of the suppression for 
central Au+Au at $\sqrtsnn$ = 200 GeV implies initial energy densities 
above 15 GeV/fm$^3$, $\sim$100 times larger than normal nuclear matter~\cite{phenix_wp}.

The equal amount of suppression for $\pi^0$ and $h^\pm$ observed above 
$p_T\approx$ 5 GeV/$c$~for the same Au+Au centrality seems to indicate
that the mechanism of quenching is independent of the identity of the 
high-$p_T$ light-quark hadron. This is expected if the suppression takes place
at the parton level {\it prior} to its 
fragmentation into a given hadron. Indeed, in this case the high-$p_T$
deficit depends only on the energy lost in the medium by the parent 
($u,d,s$) quark or gluon and not on the nature of the final leading hadron 
which will be produced with the same {\it universal} probabilities (fragmentation 
functions) which govern hadron production in the vacuum in more elementary systems.
The partons involved in high-$p_T$ hadroproduction considered in this work 
have typical momenta $\gtrsim$ 5 GeV/$c$, ten times larger than the ``bulk'' 
average momenta $\mean{p_T}\approx$ 0.55 GeV/$c$ of the system~\cite{ppg026}. 
Such energetic partons are then supposed to traverse (and lose energy in) 
the medium and hadronize in the {\it vacuum} a few tens of fm/$c$ later~\cite{wang03}.
The equal suppression of $h^\pm$ and $\pi^0$ does not by itself provide 
a conclusive argument for parton energy loss {\it before} fragmentation 
in the vacuum because above $p_T\approx$ 5 GeV/$c$, unidentified charged 
hadron yields are dominated by $\pi^\pm$~\cite{phenix_hipt_200}.
Measurement of the yields of an additional light-quark species like 
the $\eta$ meson at large enough $p_T$ allows a confirmation of 
the independence of the quenching with respect to the nature of the produced 
hadron, and tests the consistency of the data with medium-induced
{\it partonic} energy loss prior to vacuum hadronization. 
Besides its interest as a {\it signal} in its own right, 
the $\eta$ meson constitutes, after the $\pi^0$, the second most important source of 
decay $e^\pm$ and $\gamma$. 
Reliable knowledge of their production cross-sections is thus
required 
in order to eliminate the {\it background} of secondary $e^\pm$ and $\gamma$
in single electron~\cite{phenix_charm}, dielectron~\cite{phenix_dielec} 
and direct $\gamma$~\cite{ppg042} measurements.

This Letter presents measurements of the $\eta$ meson by the PHENIX experiment~\cite{nim_phenix}
in Au+Au collisions at $\sqrt{s_{_{NN}}}$ = 200~GeV during the second RHIC run 
(2001--2002) and compares them to $\eta$ from p+p and d+Au~\cite{ppg055} 
and to $\pi^0$~\cite{ppg014,ppg054} and direct $\gamma$~\cite{ppg042} from Au+Au,
all measured in the same experiment at the same $\sqrtsnn$. The $\eta$ measurement reaches the second largest 
$p_T$ for identified hadrons at RHIC, after the $\pi^0$.
The analysis reported here uses Beam-Beam Counters (BBC, $3.0<|\eta|<3.9$) 
plus the Zero Degree Calorimeters (ZDC) for trigger and global event characterization. 
For each collision, the reaction centrality is obtained by cuts in the correlated
distribution of the charge detected in the BBC and the energy measured in the ZDC~\cite{ppg001}.
A Glauber Monte Carlo model combined with a simulation of BBC and ZDC
responses is used to determine the corresponding nuclear overlap function 
$\langle T_{AA} \rangle$ 
for each centrality~\cite{ppg014}.
The $\eta$ mesons are reconstructed at mid-rapidity in the 
lead-scintillator (PbSc) electromagnetic calorimeter~\cite{nim_emc} via their 
$\gamma\gamma$ decay mode (BR = 39.43\%). The PbSc
consists of 15,552 individual lead-scintillator sandwich modules 
(5.54~cm $\times$ 5.54~cm $\times$ 37.5~cm, 18~$X_0$), grouped in six sectors
located at a radial distance of 5.1\,m from the beam line, covering a total 
solid angle of $\Delta\eta \approx 0.7$ and $\Delta \phi \approx 3\pi/4$ rad.
The energy calibration of the PbSc modules is obtained from the 
beam-test values and confirmed with the measured position of the $\pi^0$ mass peak, 
the energy deposited by minimum ionizing particles traversing the calorimeter, 
as well as with the expected $E_{\ensuremath{PbSc}}/p_{\ensuremath{tracking}}\sim$1 
value for $e^\pm$ identified by the Ring-Imaging \v{C}erenkov detector. 
The systematic error on the absolute energy scale is less than 
1.5\%, which translates into a maximum 8\% uncertainty in the final $\eta$ yields.

\begin{figure}[tbp]
\includegraphics[width=1.0\linewidth]{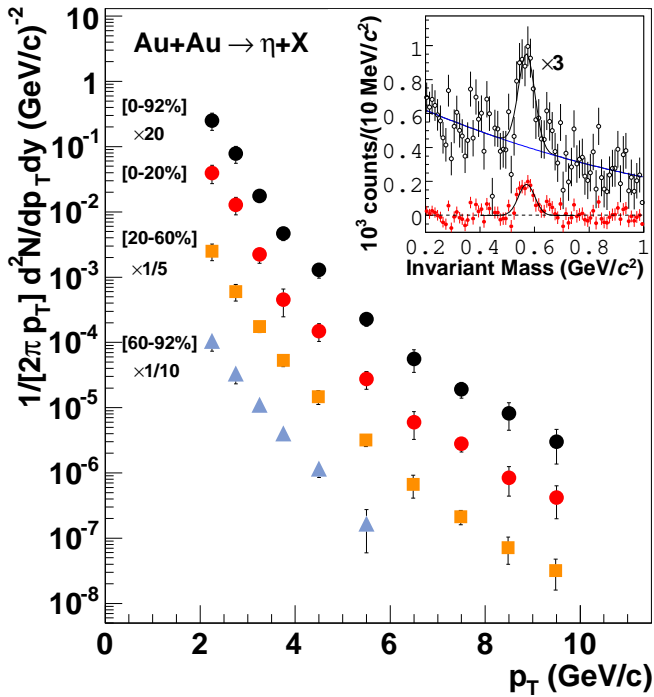}
\caption{Invariant $\eta$ yields as a function of transverse momentum for 3 centralities 
and MB Au+Au at $\sqrt{s_{_{NN}}}$~=~200~GeV scaled by the factors indicated in the plot.
Inset: Invariant mass distribution of $\gamma$ pairs with $p_T$ = 4 -- 5 GeV/$c$
measured in MB Au+Au, after mixed-event (black open circles)
plus residual (red solid circles) background subtraction.}
\label{fig:eta_spectra}
\end{figure}

For this analysis a minimum bias (MB) trigger sample of $34 \times 10^6$ events, 
also used for the previously published $\pi^0$ analysis~\cite{ppg014}, 
is combined with a Level-2 trigger event sample equivalent to an additional 
$30 \times 10^6$ minimum bias events~\cite{ppg054}. The Level-2 trigger sample is 
obtained with a software trigger on highly energetic particles (3.5 GeV threshold).
The resulting trigger reaches a 50\% (100\%) efficiency for $\eta$ 
above $p_T = 5\,(7)\;\mbox{GeV}/c$. The normalization of the Level-2 
data sample relative to the MB data sample is accurate to $2\%$.
Both sets of events are required to have a vertex position $|z|<$ 30~cm 
along the beam axis. Photon candidates are identified in the PbSc by applying
particle identification (PID) cuts based on the time-of-flight and 
shower profile~\cite{ppg014,ppg055}. The systematic uncertainty on the yields 
related to the applied PID cuts is $\sim$8\%.
The $\eta$ yields are determined by an invariant mass analysis of photon pairs
with asymmetries $|E_{\gamma 1}-E_{\gamma 2}|/(E_{\gamma 1}+E_{\gamma 2})<$ 0.5. 
The combinatorial background is obtained by combining uncorrelated photon pairs 
from different events with similar centrality and vertex, and by normalizing the 
distribution in a region below  ($m_{inv}$ = 400 -- 450 MeV/$c^2$ ) and above 
($m_{inv}$ = 750 -- 1000 MeV/$c^2$) the $\eta$ mass peak.
The resulting distribution is fit to a Gaussian plus an exponential to
account for the residual background not described by the mixed event background
(inset of Fig.~\ref{fig:eta_spectra}). The open (solid) symbols depict the $\eta$ 
signal after mixed (plus residual) background subtraction. To estimate the 
uncertainty in the subtraction procedure, different pair asymmetries and an 
alternative linear parametrization of the residual background are used. 
The signal-to-background ratio in peripheral (central) collisions is approximately 
1.3 (1.5) for the highest $p_T$ and 0.05 (0.002) for the lowest $p_T$. 

The raw spectra are normalized to one unit of rapidity and full azimuth. 
This purely geometrical acceptance factor amounts to $\sim$4 at large $p_T$.
The spectra are further corrected for the detector response (energy resolution, 
dead areas), the reconstruction efficiency (analysis cuts), and 
occupancy effects (cluster overlaps). These corrections are
quantified by embedding simulated single $\eta$ from a full 
PHENIX GEANT~\cite{geant} simulation into real events, and analyzing the merged 
events with the same analysis cuts used to obtain the real yields. 
The total $\eta$ yield efficiency correction is $\sim$3 and rises $\lesssim$20\%
with centrality. The losses are dominated by fiducial and asymmetry cuts. 
The nominal energy resolution is adjusted in the simulation by adding 
a $p_T$-independent energy smearing of $3\%$ for each PbSc tower. 
The shape, position, and width of the $\eta$ peak measured for all $p_T$'s
and centralities are well reproduced by the embedded data.


The main sources of systematic errors in the measurement are the uncertainties 
in the yield extraction (10--30\%), the yield correction (10\%), 
and the energy scale (a maximum of 8\%). The final combined 
systematic errors on the spectra are at the level of 10-30\%
(point-to-point) and 10-20\% ($p_T$-correlated) depending on the $p_T$ 
and centrality bin~\cite{ppg055}. A correction in the yield to account for 
the true mean value of each $p_T$ bin is applied to the steeply falling spectra. 
The fully corrected $p_T$ distributions are shown in Fig.~\ref{fig:eta_spectra} 
for MB and 3 centrality bins (0--20\%, 20--60\% and 60--92\%) 
scaled for clarity by the factors indicated. The error bars are the
quadratic sum of statistical and systematic errors.

\begin{figure}[tbp]
\begin{center}
\includegraphics[width=1.0\linewidth]{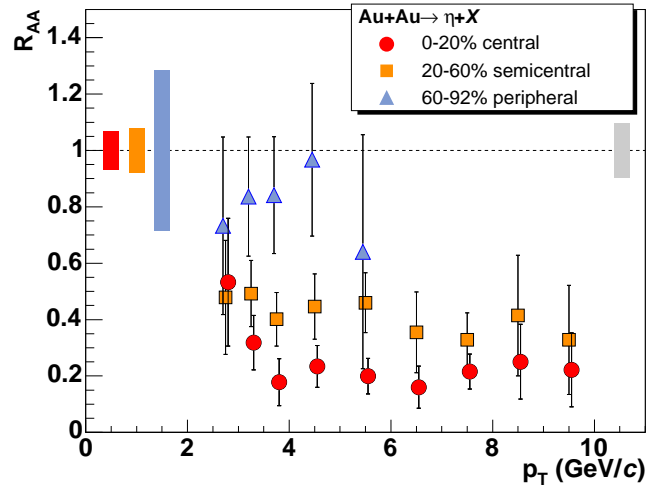}
\end{center}
\caption{Nuclear modification factors for $\eta$ in Au+Au centralities:
0--20\%, 20--60\%, 60--92\%. The error bars show point-to-point uncertainties.
The absolute normalization error bands at $R_{AA}$ = 1 show 
the uncertainties in $\mean{T_{AA}}$ for decreasing centralities. 
The error box on the right indicates the 9.7\% p+p cross-section 
uncertainty~\protect\cite{ppg044}.}
\label{fig:R_AA_eta}
\end{figure}

Medium effects in A+A collisions are quantitatively determined using
the \emph{nuclear modification factor} given as the ratio of the measured 
A+A invariant yield over the $p+p$ cross-section scaled by the
Glauber nuclear overlap function $\langle T_{AA} \rangle$
in the centrality bin under consideration:
\begin{equation}
R_{AA}(p_T)\,=\,\frac{d^2N_{AA}/dp_T dy}{\langle T_{AA}\rangle \,\cdot\,d^2\sigma_{pp}/dp_T dy}.
\label{eq:R_AA}
\end{equation}
Deviations from $R_{AA}(p_T)$ = 1 quantify the degree of departure 
of the hard A+A yields from an incoherent superposition of $NN$ collisions. 
Figure~\ref{fig:R_AA_eta} compares the nuclear modification factor for
$\eta$ in central (0--20\%), semi-central (20--60\%) and peripheral (60--92\%) 
Au+Au reactions using the reference $d^2\sigma_{pp}/dp_T dy$ spectrum 
measured in p+p collisions~\cite{ppg055}. As observed for high-$p_T$ 
$\pi^0$~\cite{ppg014,ppg054}, the $\eta$ yields are consistent with the 
expectation of independent $NN$ scatterings in peripheral reactions 
($R_{AA}\approx$ 1) but are increasingly reduced for smaller centralities. 
The $p_T$ dependence of $R_{AA}$ is flat above 4 GeV/$c$ 
as seen also for the $\pi^0$.


\begin{figure}[tbp]
\includegraphics[width=1.0\linewidth]{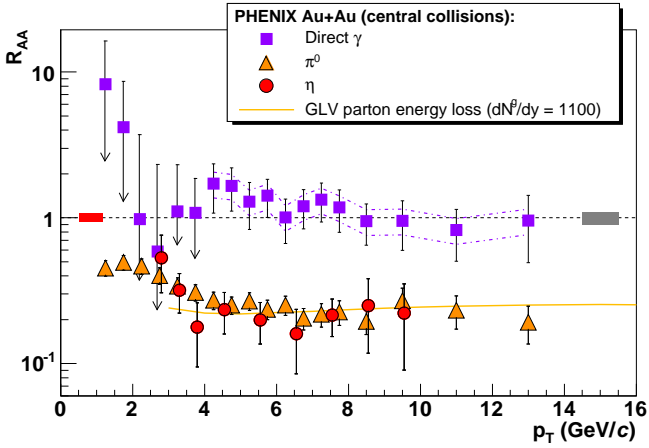}
\caption{$R_{AA}(p_T)$ measured in central Au+Au at $\sqrtsnn$~=~200 GeV for 
$\eta$, $\pi^0$~\protect\cite{ppg014,ppg054} and direct $\gamma$~\protect\cite{ppg042}. 
The error bars include all point-to-point 
errors. The error bands at $R_{AA}$ = 1 
have the same meaning as in Fig.~\ref{fig:R_AA_eta}.
The baseline p+p $\rightarrow\gamma+X$ reference used is a NLO calculation~\protect\cite{ppg042,vogelsang}, 
that reproduces our own data well~\protect\cite{ppg049}, with theoretical uncertainties 
indicated by the dash-dotted lines around the points. The solid yellow curve
is a parton energy loss prediction for a medium with density
$dN^g/dy=$ 1100~\protect\cite{vitev_gyulassy}.}
\label{fig:R_AA_eta_pi0_gamma} 
\end{figure}

Figure~\ref{fig:R_AA_eta_pi0_gamma} compares the $R_{AA}(p_T)$ 
measured in Au+Au at $\sqrtsnn$ = 200 GeV for $\eta$ (0--20\% centrality), 
$\pi^0$~\cite{ppg014,ppg054} and $\gamma$~\cite{ppg042} (0--10\% centralities).
Whereas direct $\gamma$ are unsuppressed compared to the $T_{AA}$-scaled
reference given here by a next-to-leading-order (NLO) calculation~\cite{ppg042,vogelsang} 
that reproduces the PHENIX p+p photon data well~\cite{ppg049}, 
$\pi^0$ and $\eta$ are suppressed by a similar factor of $\sim$5 
compared to the corresponding p+p cross-sections~\cite{ppg054,ppg055}.
Within the current uncertainties, light-quark mesons at RHIC 
show a flat suppression in the range $p_T$ = 4 -- 14 GeV/$c$, independent 
of their mass (note that the $\eta$ is four times heavier than the $\pi^0$). 
The results are in agreement with expectations of in-medium non-Abelian 
energy loss of the parent parton prior to its fragmentation in the vacuum. 
The initial gluon densities needed to quench the high-$p_T$ hadrons by such an amount 
are of the order of $dN^g/dy$ = 1100 (solid curve in Fig.~\ref{fig:R_AA_eta_pi0_gamma})~\cite{vitev_gyulassy}.

An additional way to determine possible differences in the suppression pattern of
$\pi^0$ and $\eta$ is to study the centrality dependence of the $\eta/\pi^0$
ratio in Au+Au collisions and compare it with the ratio in more elementary 
systems ($e^+e^-$, p+p, d+Au).
The $\eta/\pi^{0}$ ratio in hadron-hadron, hadron-nucleus and nucleus-nucleus 
collisions is seen to increase rapidly with $p_T$ and flatten out above $p_T\approx$ 2.5 GeV/$c$ 
at an asymptotically constant $R_{\eta/\pi^{0}}\approx$ 0.5 for all systems~\cite{ppg055}. 
Likewise, in $e^+e^-$ at the $Z$ pole ($\sqrt{s}$ = 91.2 GeV) 
one also finds $R_{\eta/\pi^{0}}\approx$ 0.5 for $\eta$ and $\pi^0$ at large scaled momenta
$x_{p} = p_{\ensuremath{\it hadron}}/p_{\ensuremath{\it beam}}\gtrsim$ 0.3 -- 0.7~\cite{ppg055}
consistent with the range of fractional momenta $\mean{z}$ relevant for
high-$p_T$ production discussed here.
It is interesting to test if this ratio is modified in any way
by final- and/or initial-state medium effects in Au+Au collisions at RHIC. 

\begin{figure}[tbp]
\includegraphics[width=1.0\linewidth]{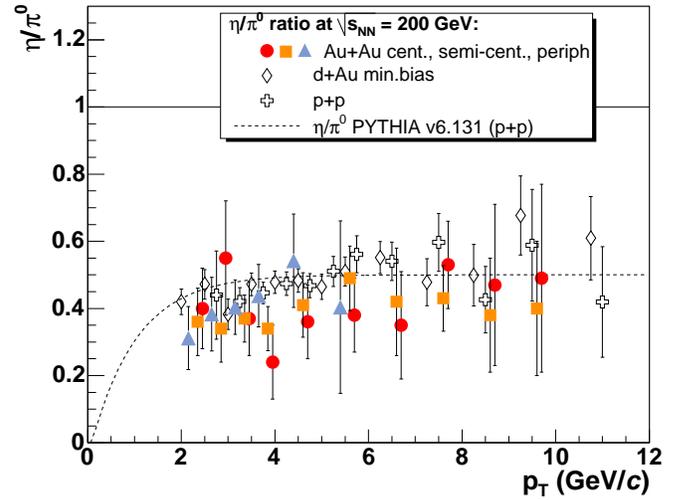}
\caption{$\eta/\pi^0$ ratio in Au+Au (centralities: 0-20\%, 20-60\%, 60-92\%) 
compared to the ratio in p+p 
and d+Au~\protect\cite{ppg055} at $\sqrt{s_{_{NN}}}$ = 200~GeV. 
The error bars include all point-to-point errors that do not cancel in 
the ratio of yields. The dashed curve is the PYTHIA~\protect\cite{pythia} 
prediction for p+p at $\sqrt{s}$ = 200 GeV consistent with the asymptotic 
$R_{\eta/\pi^{0}}\approx$ 0.5 measured in hadronic and $e^+e^-$ collisions
in a wide range of c.m. energies~\protect\cite{ppg055}.}
\label{fig:eta_pi0_ratio} 
\end{figure}

Figure~\ref{fig:eta_pi0_ratio} shows $R_{\eta/\pi^{0}}(p_T)$ for three Au+Au 
centrality selections and for p+p and d+Au collisions~\cite{ppg055}.
A fit to a constant for $p_T >$ 2 GeV/$c$ gives $R_{\eta/\pi^{0}}^{AuAu\,0-20\%}$ = 0.40 $\pm$ 0.04, 
$R_{\eta/\pi^{0}}^{dAu\,MB}$ = 0.47 $\pm$ 0.03 and $R_{\eta/\pi^{0}}^{pp}$ = 0.48 $\pm$ 0.03,
where the quoted errors are the quadratic sum of statistical and systematic uncertainties.
The Au+Au ratio is consistent within $\sim 1\sigma$ with both the essentially identical 
d+Au and p+p ratios. The $R_{\eta/\pi^{0}}$ ratio shows thus no apparent collision system, 
centrality, or $p_T$ dependence. The dotted curve is the predicted PYTHIA~\cite{pythia} 
result for the p+p ratio at $\sqrt{s}$ = 200 GeV which is also coincident with the world 
data measured in the same momentum range in hadronic, nuclear, and $e^+e^-$ collisions
in a wide range of energies ($\sqrt{s}\approx$ 3 -- 1800 GeV)~\cite{ppg055}. 

%
In summary, the transverse momentum spectra of $\eta$ mesons have
been measured at mid-rapidity in the range $p_T$ = 2--10~GeV$/c$ 
in Au+Au at $\sqrt{s_{_{NN}}}$ = 200~GeV.   The invariant yields per 
nucleon-nucleon collision are increasingly depleted 
with centrality in comparison to p+p results at the same center-of-mass 
energy. The maximum suppression factor is $\sim$5 in central Au+Au.
The magnitude, $p_T$, and centrality dependences of the suppression are 
the same for $\eta$ and $\pi^0$ suggesting that 
the production of light neutral mesons at large $p_T$ in nuclear collisions 
at RHIC is affected by the medium in the same way.
The measured $\eta$/$\pi^0$ ratio is flat with $p_T$ and amounts to 
$R_{\eta/\pi^{0}}$ = 0.40 $\pm$ 0.04. This value is consistent with the 
world value at high-$p_T$ in hadronic and nuclear reactions and, at high $x_p$, 
in $e^+e^-$ collisions. We conclude that all these observations are in 
agreement with a scenario where the parent parton first loses energy in 
the produced dense medium and then fragments into a leading meson in the 
vacuum according to the same probabilities that govern high-$p_T$ 
hadroproduction in more elementary systems (p+p, $e^+e^-$).

%


We thank the staff of the Collider-Accelerator and Physics
Departments at BNL for their vital contributions.  We acknowledge
support from the Department of Energy and NSF (U.S.A.), 
MEXT and JSPS (Japan), CNPq and FAPESP (Brazil), NSFC (China), 
CNRS-IN2P3 and CEA (France), 
BMBF, DAAD, and AvH (Germany), 
OTKA (Hungary), DAE and DST (India), ISF (Israel), 
KRF and CHEP (Korea), RMIST, RAS, and RMAE (Russia), 
VR and KAW (Sweden), U.S. CRDF for the FSU, 
US-Hungarian NSF-OTKA-MTA, and US-Israel BSF.



\end{document}